# Bipolar conduction asymmetries lead to ultra-high thermoelectric power factor

*Patrizio Graziosi,*[1,2] *Zhen Li*[1]*, Neophytos Neophytou*[1]

[1] School of Engineering, University of Warwick, Coventry, CV4 7AL, UK

[2] Consiglio Nazionale delle Ricerche – Istituto per lo Studio dei Materiali Nanostrutturati, CNR – ISMN, via Gobetti 101, 40129, Bologna, Italy

ABSTRACT.

Low bandgap thermoelectric materials suffer from bipolar effects at high temperatures, with increased electronic thermal conductivity and reduced Seebeck coefficient, leading to reduced power factor and low *ZT* figure of merit. In this work we show that the presence of strong transport asymmetries between the conduction and valence bands can allow high phonon-limited electronic conductivity at finite Seebeck coefficient values, leading to largely enhanced power factors. The power factors that can be achieved can be significantly larger compared to their maximum unipolar counterparts, allowing for doubling of the *ZT* figure of merit. We identify this behavior in low-bandgap cases from the half-Heusler material family. Using both, advanced electronic Boltzmann transport calculations for realistic material bandstructures, as well as model parabolic electronic bands, we elaborate on the parameters that determine this effect. We then develop a series of descriptors which can guide machine learning studies in identifying such classes of materials with extraordinary power factors at nearly undoped conditions. For this we test more than 3000 analytical bandstructures and their features, and more than 120 possible descriptors, to identify the most promising ones that contain: i) only bandstructure features for easy identification from material databases, and ii) bandstructure and transport parameters that provide much higher correlations, but for which parameter availability can be somewhat more scarce.

Thermoelectric materials convert heat from temperature gradients directly into electricity and vice-versa, and they can be used for clean energy production and solid-state heat-pumps. Thus, they can contribute to energy sustainability and reduction in the use of fossil fuels. [1] Despite their enormous potential, their large-scale exploitation is hampered by material high costs and toxicity. Large experimental and computational efforts are undertaken to discover new, more efficient and/or non-toxic thermoelectric (TE) materials. [1-10]

The TE conversion efficiency is quantified by the dimensionless figure of merit $ZT = \frac{\sigma S^2}{\kappa_L + \kappa_e} T$, where $\sigma$ is the electrical conductivity, $S$ the Seebeck coefficient, $\kappa_L$ and $\kappa_e$ are the lattice and electronic thermal conductivities, respectively, and $T$ is the absolute temperature. The $\sigma S^2$ term is called Power Factor (PF) and is responsible for the power output of the TE generator. [11-16]

High doping and a sizable material bandgap are beneficial for TEs, as the general strategy is to avoid bipolar conduction which degrades $S$ and increases $\kappa_e$. However, we have recently shown that narrow bandgap materials can offer the unconventional possibility to achieve extremely high power factors when undoped, under the condition of highly asymmetric bandstructure or transport features between the conduction and valence bands. [17] This theoretically predicted observation could interpret three recent experimental observations as well. [17-20] Specifically, the experimental work of reference [20] appeared after the publication of our original work and provides another indication of our predictions for high PF performance in ScNiSb. [20] In that work, it is also claimed that asymmetry in the electron and hole mobility and the narrow bandgap bipolar effects are responsible for high PFs.

In this letter we identify which properties determine whether a material can display this unconventional effect and develop appropriate descriptors. Our purpose is to provide understanding for experimentals, but also enable the discovery of such novel TE materials through automated, machine learning techniques.

We first describe the mechanism behind this ultra-high PF effect with reference to our original paper. [17] We consider a prototypical isotropic parabolic bandstructure with a relatively small bandgap (regardless of the bandgap being direct or indirect) and with very asymmetric conduction and valence bands (CB and VB) in terms of effective mass and charge scattering, the latter exemplified by differences in the deformation potentials (**Figure 1a**). The transport calculations are based on the Boltzmann Transport Equation, BTE, including electron-phonon (e-ph) plus electron ionized impurity scattering (IIS) as detailed in Ref. [1**7**]. With regards to IIS, we consider the generic $\partial n/\partial E_F$ term in the calculation of the screening length, which allows for capturing the behavior from the weak to the strong screening regimes and in-between, which is important for our results.

From the plot of the conductivity versus the Fermi level position, $\eta_F$, (**Figure 1b**), the conductivity (**blue line**) features a peak at the intrinsic Fermi level, at which we set the reference $\eta_F = 0$, where it reaches the phonon-limited value (**red line**). Around the intrinsic region the electron and hole densities are similar, $n = p$, making the underlying dopant impurity density negligible. This makes IIS having no effect on transport and the conductivity to become phonon-limited. However, the individual $n$, $p$ concentrations can be significant and result in high conductivity, as carriers can be thermally excited to the CB from the VB over the narrow bandgap. We find that this effect is significant for bandgaps $< 6k_BT$. The second necessary condition is related to the Seebeck coefficient, $S$. For symmetric bands with comparable electron and hole mobilities, $S$ is close to zero in the intrinsic region; more specifically $S = 0$ when $n\mu_n = p\mu_p$, with $\mu_{n/p}$ being the electron/hole mobilities. However, highly asymmetric mobilities arising from very asymmetric scattering rates due to very different CB/VB density of states (DOS) and deformation potentials, separates the $S = 0$ and the $\eta_F = 0$ energy points (inset of **Figure 1b**). The finite $S$ value at $\eta_F = 0$ allows for the conductivity peak to be transferred in the PF (**Figure 1c**). On the other hand, since $\kappa_e$ also peaks at $\eta_F = 0$, [17] the impact on $ZT$ is mitigated to around 2× (see **Figure 1d**), which is still, however, significant and it is essentially the same improvement as what nanostructuring provided in the wider materials $ZT$ range.[12] We basically show that there is another way to achieve this improvement using highly asymmetric band materials. We stress here that this effect originates from achieving phonon-limited transport conditions, i.e. the red dash-dot lines in **Figures 1b-d**, and appears when all the other scattering mechanisms are weaker than electron-phonon scattering at $\eta_F = 0$, which indicates the intrinsic Fermi level position.

Importantly, we also observe this effect in full-band calculations for some half-Heusler compounds, where we considered energy and momentum dependent scattering rates[21] and full bipolar considerations[22] by using the *ElecTra* simulator.[23] Simulations of four narrow gap compounds featuring ultra-high PF peaks in their intrinsic region around $\eta_F = 0$ are presented in **Figure 2a, c** for $T = 300$ K and 900 K, respectively. Note that the doping range in which this effect is observed at, is much wider, as shown in **Figure 2b, d**. The PF peak is pronounced even more at higher temperatures as bipolar transport increases, (**Figures 2c, d**). Also, materials of somewhat larger bandgap, which do not show the peak at 300 K, can develop this behavior at higher temperatures. Such PF increase with temperature results from the $E_g/k_B T$ reduction, which drives the material more into the bipolar regime, in contrast to the common unipolar operation which mostly experiences PF reduction with temperature.[21] Note that for the 300 K and 900 K cases we left the bandgap unchanged, however we later on present a study of using the bandgap as a free parameter.

We also note that under these asymmetric bandstructure and transport conditions, the polarity of the pristine materials, and hence the PF peak, is dictated by the polarity of the carriers with the highest mobility. Thus, a material features *n*-type transport even when (lightly) doped with acceptor impurities (HfNiSn and NbFeSb in **Figure 2**), or vice-versa features *p*-type transport when (lightly) doped with donors (Sc- based in **Figure 2**).

We now focus on the identification of material parameters and descriptors which better correlate with the high PF effect. For this we consider parabolic bands and keep the VB parameters fixed, with DOS and conductivity masses ($m_{DOS}$ and $m_{cond}$) equal to the electron rest mass $m_0$. We consider acoustic deformation potential ($D_{ADP}$) of 10 eV, and a mass density and sound velocity are as in the caption of **Figure 1**. We then vary several CB parameters as depicted in **Figure 3a**: we vary the $m_{DOS}$ from 0.2 to $0.9m_0$, the $m_{cond}$ from $0.25m_{DOS}$ (strongly anisotropic) to $1m_{DOS}$ (isotropic valley), the $D_{ADP}$ from 1 to 9 eV, and the bandgap from 0.2 to 0.6 eV. A total of 3168 distinct possibilities are formed. With these inputs, the polarity under pristine conditions is always *n*-type, but the outcomes can be extended to the VB mutatis mutanda. We performed the calculation for three temperatures, 300 K, 600 K, and 900 K. Based on the underlying transport physics,[17] we consider seven possible important material parameter combinations that can form descriptors, as:

$$\frac{|m_{\text{cond}}^{min}-m_{\text{cond}}^{maj}|}{m_{\text{cond}}^{min}+m_{\text{cond}}^{maj}} \quad \text{1a}$$

$$\frac{|m_{\text{DOS}}^{min}-m_{\text{DOS}}^{maj}|}{m_{\text{DOS}}^{min}+m_{\text{DOS}}^{maj}} \quad \text{1b}$$

$$\frac{|D^{min}-D^{maj}|}{D^{min}+D^{maj}} \quad \text{1c}$$

$$e^{-E_g/k_B T} \quad \text{1d}$$

$$\left(\frac{m_{\text{DOS}}^{min}}{m_{\text{DOS}}^{maj}}\right)^{\frac{3}{2}} \quad \text{1e}$$

$$(E_g/k_B T)^2 \quad \text{1f}$$

$$\frac{1}{D^{maj^2} m_{\text{cond}}^{maj}} \quad \text{1g}$$

In the terms above, $D$ is the deformation potential (we consider only ADP and IIS), $m$ the effective mass and $E_g$ the bandgap, and subscripts '*maj*' and '*min*' refer to majority and minority carriers. Terms 1a to 1c express the asymmetry in the conductivity mass, the DOS mass, and the e-ph scattering strength, respectively. Term 1d captures the excited minority carriers. Term 1e captures the stabilization of the intrinsic Fermi level close to the CB or VB. Term 1f indicates that a large bandgap is beneficial for *S*. Term 1g captures a high phonon-limited PF, since the high PF value in the intrinsic region is due to the fact that the ph-limited transport is reached, we want a weak e-ph coupling (low deformation potential) and small conductivity effective mass .[11] Notably, terms 1a,b,d,e,f can be obtained from DFT bandstructure calculations alone. The effective masses of interest can be extracted from a dedicated code that we have developed,[24] and $E_g$ can be extracted by DFT with a careful choice of pseudopotentials. [25-28] Terms 1c and 1g require deformation potentials , which can be extracted by *ab initio* calculations. [29,30]

Combinations of these seven terms lead to 127 possible descriptors. For these we show in **Figure 3b** the Pearson correlation coefficients, *r*, of the 3168 distinct bandstructures for each of the three temperatures we consider (blue, orange, and yellow symbols for 300, 600 and 900 K, respectively). The green circles highlight the most relevant and promising descriptors which are:

$$\frac{|m_{cond}^{min}-m_{cond}^{maj}|}{m_{cond}^{min}+m_{cond}^{maj}} \frac{|m_{DOS}^{min}-m_{DOS}^{maj}|}{m_{DOS}^{min}+m_{DOS}^{maj}} e^{-E_g/k_BT} (E_g/k_BT)^2 \qquad \text{2a}$$

$$\frac{1}{D^{maj^2} m_{cond}^{maj}} \qquad \text{2b}$$

$$e^{-E_g/k_BT} (E_g/k_BT)^2 \frac{1}{D^{maj^2} m_{cond}^{maj}} \qquad \text{2c}$$

$$\frac{|m_{cond}^{min}-m_{cond}^{maj}|}{m_{cond}^{min}+m_{cond}^{maj}} e^{-E_g/k_BT} (E_g/k_BT)^2 \frac{1}{D^{maj^2} m_{cond}^{maj}} \qquad \text{2d}$$

$$\frac{|D^{min}-D^{maj}|}{D^{min}+D^{maj}} e^{-E_g/k_BT} (E_g/k_BT)^2 \frac{1}{D^{maj^2} m_{cond}^{maj}} \qquad \text{2e}$$

$$\frac{|m_{cond}^{min}-m_{cond}^{maj}|}{m_{cond}^{min}+m_{cond}^{maj}} \frac{|D^{min}-D^{maj}|}{D^{min}+D^{maj}} e^{-E_g/k_BT} (E_g/k_BT)^2 \frac{1}{D^{maj^2} m_{cond}^{maj}} \qquad \text{2f}$$

$$\frac{|m_{DOS}^{min}-m_{DOS}^{maj}|}{m_{DOS}^{min}+m_{DOS}^{maj}} \frac{|D^{min}-D^{maj}|}{D^{min}+D^{maj}} e^{-E_g/k_BT} (E_g/k_BT)^2 \frac{1}{D^{maj^2} m_{cond}^{maj}} \qquad \text{2g}$$

$$\frac{|m_{cond}^{min}-m_{cond}^{maj}|}{m_{cond}^{min}+m_{cond}^{maj}} \frac{|m_{DOS}^{min}-m_{DOS}^{maj}|}{m_{DOS}^{min}+m_{DOS}^{maj}} \frac{|D^{min}-D^{maj}|}{D^{min}+D^{maj}} e^{-E_g/k_BT} \left(\frac{m_{DOS}^{min}}{m_{DOS}^{maj}}\right)^{\frac{3}{2}} (E_g/k_BT)^2 \frac{1}{D^{maj^2} m_{cond}^{maj}} \qquad \text{2h}$$

Expression 2a is the best performing descriptor among the ones which contain only DFT bandstructure terms. It provides low, but positive correlation. The descriptor in expression 2b dictates the high phonon limited PF, [22] which is reached in the pristine region, and is of sufficiently high *r*. The other six descriptors, from the simplest to the most complex also indicate high *r* values. The descriptors are basically composed of three elements which capture: i) asymmetry in the DOS ($A_{DOS}$), which brings the intrinsic Fermi level close to one band; ii) asymmetry in the transport parameters ($A_{tr}$), which separates the $S = 0$ and the $\eta_F = 0$ points; and iii) a part related to the bandgap ($A_{gap}$). We seek to maximize $A_{DOS} \times A_{tr} \times A_{gap}$, where the last two terms seem to be the most important ones.

The temperature-dependent performance of these eight descriptors is displayed in **Figure 4a-c**, for the 3168 parabolic bands (**dash-dot lines**) and 9 half-Heusler materials, 4 of which feature the high PF peak in the pristine regime (**solid lines**). The labels correspond to expression 2. **Figure 4a** reports the descriptors 2a and 2b, respectively. 2a is best one with terms only from DFT and 2b the best one (ph-limited) under the usual unipolar transport considerations. More complete and efficient descriptors are depicted in **Figure 4b** and the most complex ones in **Figure 4c**, which include full asymmetries of the DOS and conductivity effective masses. For the half-Heusler alloys we used the dominant deformation potential, which turns out to be the optical one.[22] All the relevant deformation potential values can be found in [22].

We observe that the descriptors identified with simple parabolic bands also provide high *r* for the half-Heuslers at low temperature, but fail at higher temperatures. The only expression which shows very high *r* for both the parabolic and realistic bands is the descriptor by expression 2h, which has the presence of the term 1e (DOS masses). The failure of the prior descriptors at higher temperatures in half-Heuslers occurs likely because the non-parabolicity and satellite valleys in the realistic bandstructures become relevant at higher temperatures. By performing the analysis of the 127 descriptors for the 9 small gap half-Heusler materials we consider, (HfNiSn, NbFeSb, ScNiBi, ScNiSb, TiNiSn, YNiBi, YNiSb, ZrNiPb, ZrNiSn),[22] we obtain the results reported in **Figure 4d**. Many combinations work well at 300 K, but many fail at 900 K. The two better performing descriptors at *all* temperatures are highlighted with **black circles**. The best one at the right-hand side of **Figure 4d** corresponds again to expression 2h, which captures as much asymmetry as possible. The second best one at the left-hand side is:

$$\frac{|D^{min}-D^{maj}|}{D^{min}+D^{maj}} \left(\frac{m_{DOS}^{min}}{m_{DOS}^{maj}}\right)^{\frac{3}{2}} \frac{1}{D^{maj^2} m_{cond}^{maj}} \qquad 3.$$

It seems that the term which stabilizes the intrinsic Fermi level close to the lighter band edge, term 1e (middle term in 3), plays a much more important role at the higher temperatures in complex bandstructures than in simple parabolic ones. However, a much larger sample of realistic bandstructures should be used to extract better conclusions, which at this point it is not available.

Note that we don't consider non-parabolicity here, but we have performed the same study with non-parabolicity in the Supporting Information (SI). There we show that the descriptors are still valid under non-parabolic band conditions, except for a slight descriptor degradation at elevated temperatures, but which can be fixed by accounting for the overall effective mass of the transport carries increasing with temperature (see SI).

Most TE materials are nanostructured. Since the condition to observe this high PF in the undoped region of narrow-gap TEs is that the e-ph scattering rate is weak compared to all the other scattering mechanisms (to reach phonon-limited transport). It is useful to obtain an order of magnitude estimate for the grain size in polycrystalline/nanostructured materials that will still allow for this effect. [12] For this, we compute the mean-free-path $\lambda_{(E,E_F,T)}$ using the *ElecTra* software [12] for each transport state, and then extract a comprehensive mean-free-path (mfp) as:

$$\text{mfp}_{(E_F,T)} = \frac{\int_E \lambda_{(E,E_F,T)} g_{(E)} f_{(E,E_F,T)}}{\int_E g_{(E)} f_{(E,E_F,T)}} \qquad (4),$$

where $g_{(E)}$ is the DOS and $f_{(E,E_F,T)}$ the Fermi-Dirac distribution. [21] The evaluated mfps for the four half-Heusler alloys with the high PFs in the intrinsic region for $T = 600$ K, are shown in **Figure 4e** with the same color scheme as in **Figure 2**. Since the material mfps are a few 100s of nm, then for weaker grain boundary scattering, grains of 1 $\mu$m or larger would be sufficient to allow high PF. NbFeSb would require somewhat larger grains, but it is well-known that micrometers size grains offer higher PF in that material. [31]

Still, however, the PF can be larger compared to the unipolar PF when the material is highly doped material, and the exact behavior needs further examination. For example, it is possible that in the bipolar regime the electronic part of the thermal conductivity $\kappa_e$ dominates the overall thermal conductivity (see inset of Fig. 1d). In that way nanostructuring could be irrelevant to the *ZT* as it will have the same effect on both $\kappa$ (through the dominant $\kappa_e$) and $\sigma$. But the overall effect of nanostructuring could be more complicated, involving the effect on $\kappa_{ph}$, and phenomena such as energy filtering. [11,32-35] For the latter, energy filtering either for majority or minority carriers would reduce bipolar effects (effectively increase the bandgap) and reduce the effect we describe. On the other hand, if enough energy space of potential barriers occupies the light highly conductive band region, as in a superlattice, then this can shift the intrinsic level even higher into the light bands and provide larger conductivity, which could boost the effect we describe. [12,36]

In conclusion, we highlighted the material parameters which allow for ultra-high PFs in the pristine undoped region of certain narrow gap bipolar materials. We offered a useful set of motivated descriptors that can enable the screening of material data sets (2h and 3). Our results show that in general we seek materials which maximize the transport asymmetry between their conduction and valence bands through i) asymmetry in the effective masses (either DOS and/or conductivity), ii) asymmetry in e-ph scattering strength, and iii) a small bandgap of a few $k_BT$.

Figure 1:

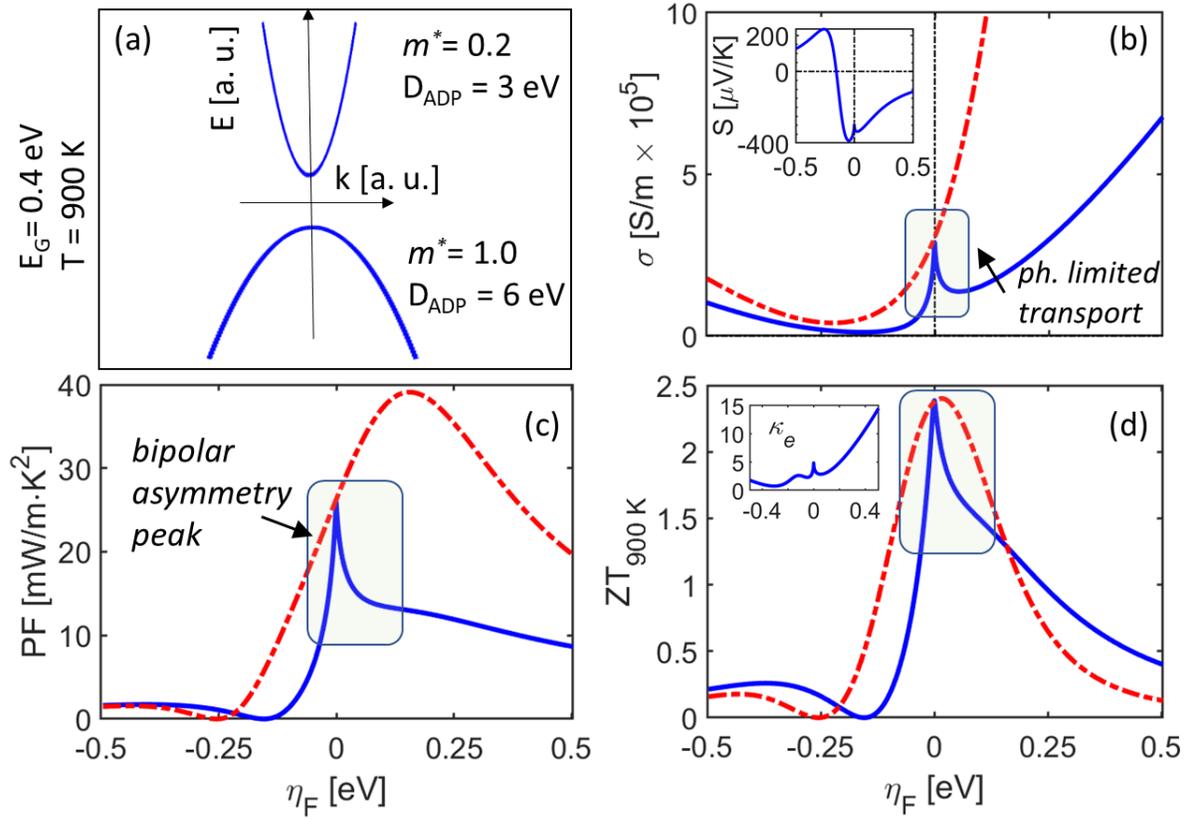

**Figure 1**: (a) Schematic of highly anisotropic parabolic bands. (b) Electrical conductivity with the parameters indicated in (a). The dielectric constant is taken to be 12, the mass density 6 Kg/m$^3$, the speed of sound 4000 m/s. Inset: Seebeck coefficient, *S*. It is noticeable that *S* is not zero at the intrinsic Fermi level ($\eta_F = 0$). (c) Power Factor, PF. (d) Figure of merit, *ZT*. Lattice thermal conductivity of 5 W/mK is used. The blue lines show phonon+IIS whereas the dash-dot red lines the phonon-limited case. Inset: electron thermal conductivity in units of W/Km.

Figure 2:

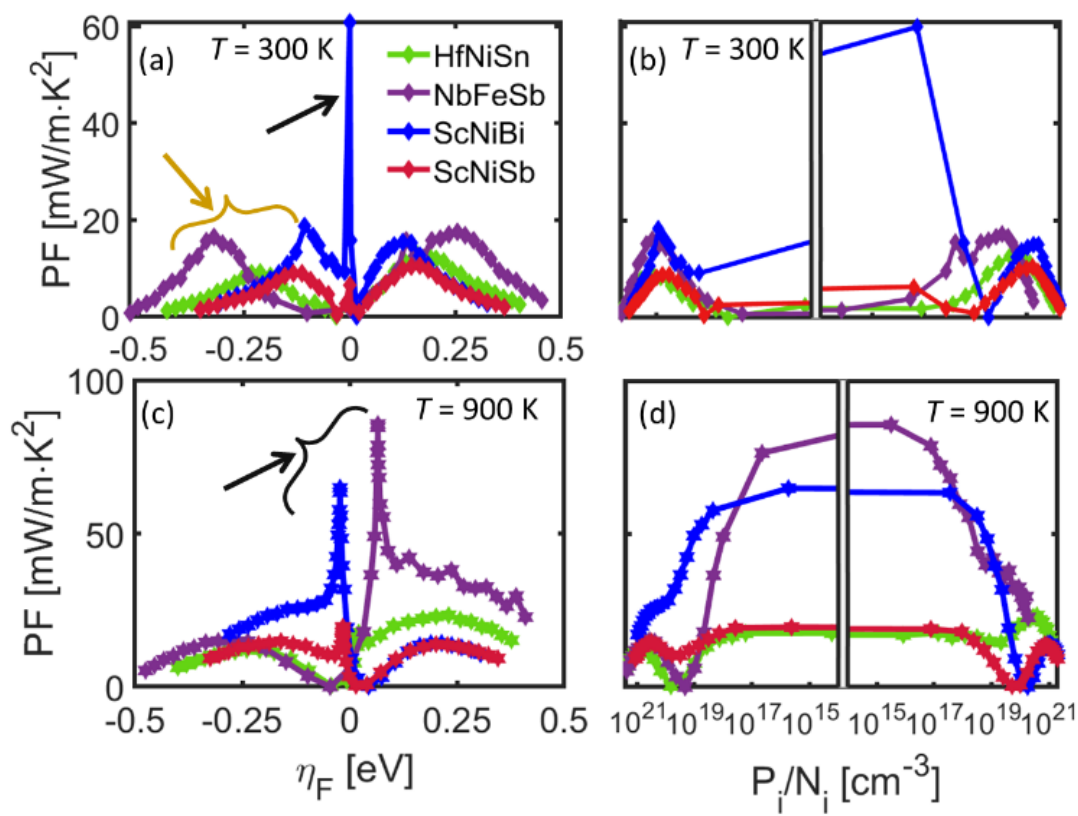

**Figure 2**: (a) PF versus Fermi level at $T = 300$ K. The gold arrow indicates the usual, unipolar PF peak. The black arrows indicate the additional ultra-high PF arising from narrow gap conditions. (b) As in (a) but versus impurity density; left-hand side: *p*-type doping, right-hand side: *n*-type doping. (c) and (d) are as (a) and (b) but for 900 K.

Figure 3:

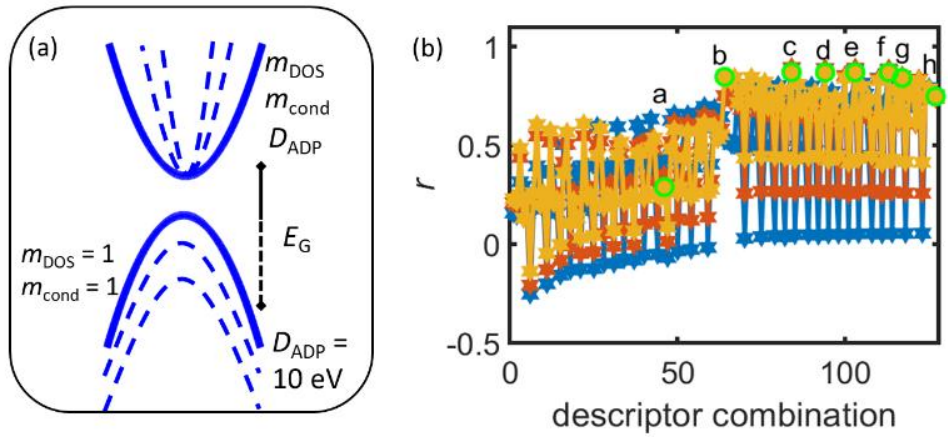

**Figure 3**: (a) Schematic of the prototypical parabolic bandstructures used for identifying descriptor performance, indicating the variation parameters, for a total of 3168 combinations. (b) Pearson correlation coefficients for the "ultra-high" PF value in the pristine regime for the 3168 combinations and three temperatures: 300 K (blue), 600 K (orange), 900 K (yellow). The descriptors highlighted with green circles are labelled with the letter which correspond to eq. 2 in the text.

Figure 4:

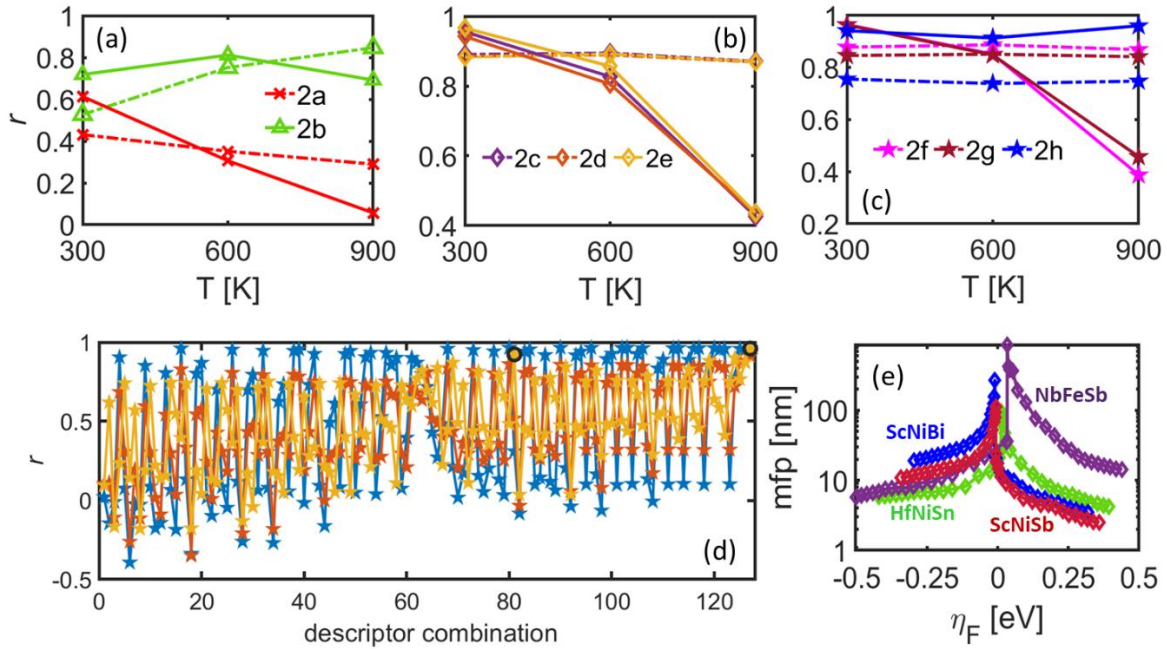

**Figure 4**: (a-c) Temperature dependence of the descriptors in eq. 2, to which the legends refer; dash-dot lines are for parabolic bands and solid ones for half-Heusler alloys. (a) Descriptor which contains only terms obtainable from DFT (2a) and for high phonon-limited transport (2b). (b) Three well-performing descriptors combining bandstructure and scattering parameters. (c) Three descriptors which capture more asymmetry. (d) The 127 descriptors, as in **Figure 3b**, for nine half-Heusler compounds. The black circles highlight two very good descriptors as noted in the text. (e) Mean-free-path (mfp) for four half-Heusler materials with high PF in the pristine region.

AUTHOR INFORMATION

**Corresponding Author**

*Patrizio.Graziosi@cnr.it, Patrizio.Graziosi@gmail.com

**Author Contributions**

The manuscript was written through contributions of all authors.

ACKNOWLEDGMENT

This work has received funding from the Marie Skłodowska-Curie Actions under the Grant agreement ID: 788465 (GENESIS - Generic semiclassical transport simulator for new generation thermoelectric materials) and from the European Research Council (ERC) under the European Union's Horizon 2020 Research and Innovation Programme (Grant Agreement No. 678763). For part of the computational time, we acknowledge the CINECA award under the ISCRA initiative, for the availability of high performance computing resources and support.

SUPPLEMENTAL MATERIALS

Online Supporting Information present the simulations for non-parabolic bands and the descriptors values for the half-Heusler compounds.

# Supporting Information

Bipolar conduction asymmetries lead to ultra-high thermoelectric power factor

*P. Graziosi, Z. Li, N. Neophytou*

## *Descriptors for non-parabolic bands*

We report here the calculations for non-parabolic bands, Kane's model, by inserting the non-parabolicity only in the CB, while the VB remains isotropic and parabolic as in the main manuscript. For the DOS and band velocity we use the following equations:

$$\text{DOS} = \frac{\sqrt{2}}{\pi^2 \hbar^3} m_{\text{DOS}}^{3/2} (1 + 2\alpha E)\sqrt{E(1 + \alpha E)} \tag{s1a}$$

$$v = \frac{\sqrt{2E/m_c}}{(1+\alpha E)^{1/2} + \alpha E(1+\alpha E)^{-1/2}} \tag{s1b}$$

Figure 1a below shows the DOS for an anisotropic non-parabolic band with $\alpha = 0.5$ eV$^{-1}$, $m_x = 0.1 m_0$, $m_y = 0.5 m_0$, $m_z = m_0$, fully numerically computed with the *ElecTra* code [DOI 10.5281/zenodo.5074943], and analytically calculated for the associated $m_{\text{DOS}} = 0.3684 m_0$; Figures 1b-c show the TDF and the PF for ADP scattering mechanism, computed for the same bandstructure with the numerical *ElecTra* code for the *xx* direction and analytically with $m_{\text{DOS}} = 0.3684 m_0$ and $m_c = 0.1 m_0$. These figures show an excellent agreement, supporting the reliability of our analytical treatment.

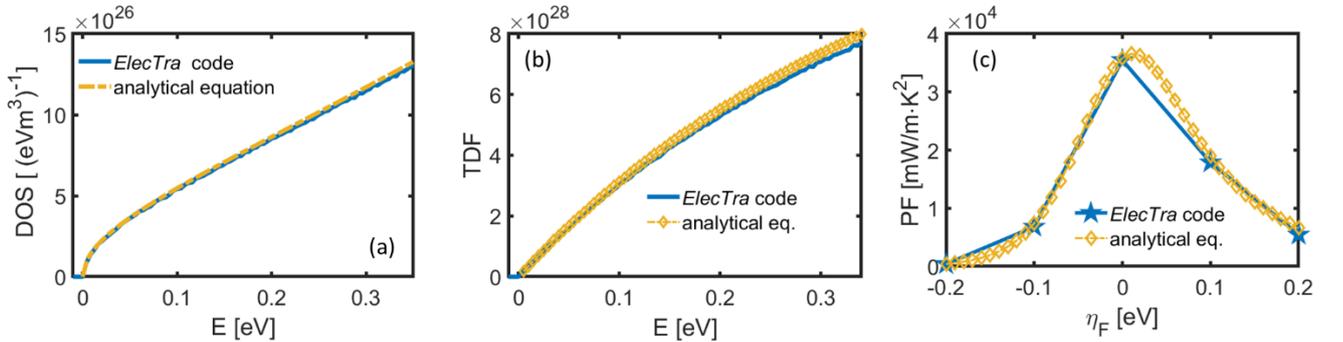

Figure S1: comparison between full numerical results, computed with the *ElecTra* code, and the one computed based on the equations 1, for an anisotropic non-parabolic band, with masses as in the text, for: (a) DOS, (b) TDF (ADP scattering), (c) PF (ADP scattering).

For the IIS scattering, we modified the Eq. proposed by Lundstrom [2] for parabolic bands, used in the manuscript in ref. [1], by varying the single effective mass value after Eq. (s1a) above, to obtain an energy dependent effective mass as:

$$m_{\text{DOS,non-parab.}}(E) = m_{\text{DOS}} \left((1 + 2\alpha E)\sqrt{1 + \alpha E}\right)^{2/3} \tag{s2}$$

For the anisotropic non-parabolic conduction bands, $\alpha$ was varied from 0 to 2 at steps of 0.5, while the other parameters were varied as in the main text, for a total of 15840 bandstructures for each temperature.

We first employ the descriptors exactly as in the text (neglecting the nonparabolicity in the descriptors). Figure 2a below highlights the best performing descriptors across the three temperatures, which are, in the order as in the figure from left to right:

$$e^{-E_g/k_BT}(E_g/k_BT)^2 \frac{1}{D^{maj^2} m_{cond}^{maj}} \quad \text{(s3a)}$$

$$\frac{|m_{cond}^{min}-m_{cond}^{maj}|}{m_{cond}^{min}+m_{cond}^{maj}} e^{-E_g/k_BT}(E_g/k_BT)^2 \frac{1}{D^{maj^2} m_{cond}^{maj}} \quad \text{(s3b)}$$

$$\frac{|D^{min}-D^{maj}|}{D^{min}+D^{maj}} e^{-E_g/k_BT}(E_g/k_BT)^2 \frac{1}{D^{maj^2} m_{cond}^{maj}} \quad \text{(s3c)}$$

$$\frac{|m_{cond}^{min}-m_{cond}^{maj}|}{m_{cond}^{min}+m_{cond}^{maj}} \frac{|D^{min}-D^{maj}|}{D^{min}+D^{maj}} e^{-E_g/k_BT}(E_g/k_BT)^2 \frac{1}{D^{maj^2} m_{cond}^{maj}} \quad \text{(s3d)}$$

labelled 2c, 2d, 2e and 2f, respectively, in the main manuscript. These are the same best descriptors identified in the manuscript for parabolic bands. Here, for the non-parabolic bands we notice a decrease in the high temperature behavior from > 0.85 to ~ 0.65, which is not observed in the parabolic case, but which was indeed observed in the full band treatment of the half-Heusler materials, in which case the bands are non-parabolic at high energies.

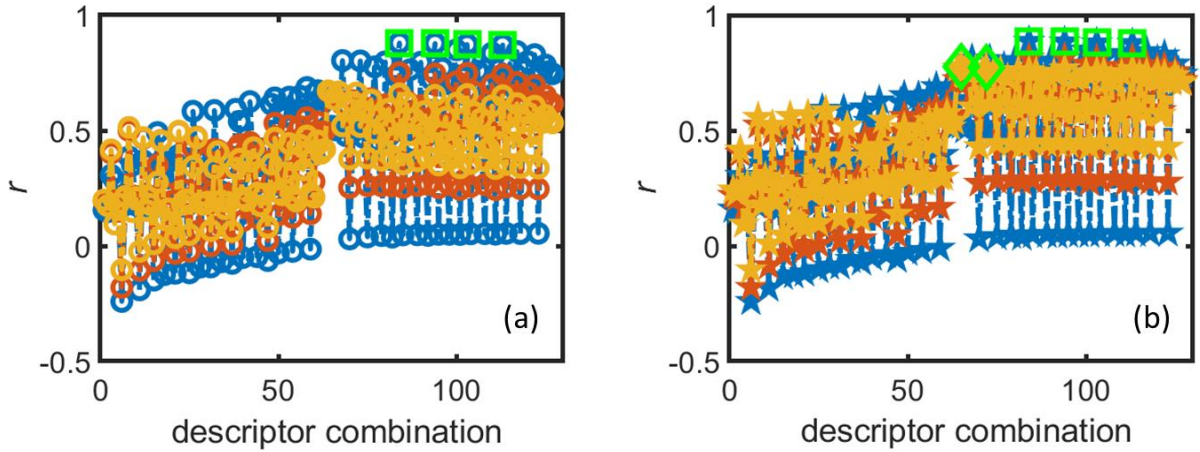

Figure S2: (a) Pearson correlation coefficients for the "ultra-high" PF value in the pristine regime for the 15840 combinations of anisotropic non parabolic bandstructures and three temperatures: 300 K (blue), 600 K (orange), 900 K (yellow), using the same descriptors as in the main manuscript. (b) the same for the case where the effective masses are multiplied by $(1 + 10\alpha k_BT)\sqrt{1 + 5\alpha k_BT}$.

Then we repeat the descriptors analysis by multiplying the majority carrier masses by $(1 + 10\alpha k_\text{B}T)\sqrt{1 + 5\alpha k_\text{B}T}$ in the individual descriptors # 1a, 1b, 1e, 1g, with reference to the manuscript labels, after considering the non-parabolic energy dependence of the DOS in equation (1a) and setting $5k_\text{B}T$ as the average transport energy window of interest. We report the results in Figure 2b, leading to the same best performing descriptors, but with a noticeable improvement for the highest temperatures, from ~ 0.65 to > 0.83.

Figure 2a highlights also with green diamonds two descriptors which performs extremely well at $T$ = 900 K, $r$ = 0.84, but less at lower temperatures. These are, left to right,

$$\frac{|m_\text{cond}^{min}-m_\text{cond}^{maj}|}{m_\text{cond}^{min}+m_\text{cond}^{maj}} \frac{1}{D^{maj^2} m_\text{cond}^{maj}} \tag{s4a}$$

$$\frac{|m_\text{cond}^{min}-m_\text{cond}^{maj}|}{m_\text{cond}^{min}+m_\text{cond}^{maj}} \frac{|D^{min}-D^{maj}|}{D^{min}+D^{maj}} \frac{1}{D^{maj^2} m_\text{cond}^{maj}} \tag{s4b}$$

These show that the anisotropy in the conductivity effective masses is the most important factor at higher temperature. The reduced role of the bandgap related term is because at 900 K the highest used bandgap, 0.6 eV, is less than ~ $8k_\text{B}T$, leading to thermally excited minority carriers at any gap value.

The results show that the proposed descriptor generally holds also for non-parabolic band even if a small decrease is observed. Importantly, non-parabolicity leads to a ≤ 25 % loss in descriptors performances at the highest temperatures, this clarifies that the decrease in the Pearson correlation value in the half-Heusler materials at $T$ = 900 K, Figure 4 in the manuscript, is likely due to non-parabolicity of the bandstructures.

*Relative descriptor values for half-Heusler materials*

The figure below shows the descriptors values computed from each small gap half-Heusler compound we consider and a couple of other semiconductors, namely silicon and germanium. The descriptors are the ones labelled 2d, 2f and 2h in the main text, respectively:

$$\frac{|m_\text{cond}^{min}-m_\text{cond}^{maj}|}{m_\text{cond}^{min}+m_\text{cond}^{maj}} e^{-E_g/k_BT} (E_g/k_\text{B}T)^2 \frac{1}{D^{maj^2} m_\text{cond}^{maj}} \tag{s5a}$$

$$\frac{|m_\text{cond}^{min}-m_\text{cond}^{maj}|}{m_\text{cond}^{min}+m_\text{cond}^{maj}} \frac{|D^{min}-D^{maj}|}{D^{min}+D^{maj}} e^{-E_g/k_BT} (E_g/k_\text{B}T)^2 \frac{1}{D^{maj^2} m_\text{cond}^{maj}} \tag{s5b}$$

$$\frac{|m_\text{cond}^{min}-m_\text{cond}^{maj}|}{m_\text{cond}^{min}+m_\text{cond}^{maj}} \frac{|m_\text{DOS}^{min}-m_\text{DOS}^{maj}|}{m_\text{DOS}^{min}+m_\text{DOS}^{maj}} \frac{|D^{min}-D^{maj}|}{D^{min}+D^{maj}} e^{-E_g/k_BT} \left(\frac{m_\text{DOS}^{min}}{m_\text{DOS}^{maj}}\right)^{\frac{3}{2}} (E_g/k_\text{B}T)^2 \frac{1}{D^{maj^2} m_\text{cond}^{maj}} \tag{s5c}$$

They are depicted in figure S3 by the blue, brown and purple bars and labelled 2d, 2f and 2h, respectively. In the panels (a) and (b), the materials are ranked by their 2f descriptor performance (brown bars). The green diamonds indicate the materials from Fig. 2 of the main paper which we analysed and have high PFs in the intrinsic region. For clarity, figures S3(a-c) show the descriptors in linear scale, whereas figure S3(d-f) in logarithmic scale. Clearly, at 300 K descriptors 2d and 2f are adequate in identifying materials with PF peaks in the intrinsic region. At 900 K many other materials with PF peaks appear with high descriptor values. 2h, seems to capture the behaviour of the best performing materials again, over-emphasizing the best performers. Evidently, Si and Ge, which do not show this bipolar PF effect, score very low in the descriptors, although most of their reduced value comes from their large bandgaps. Figures S3(d-f) show the same data in logarithmic scale for reference, indicating the low values that Si and Ge obtain.

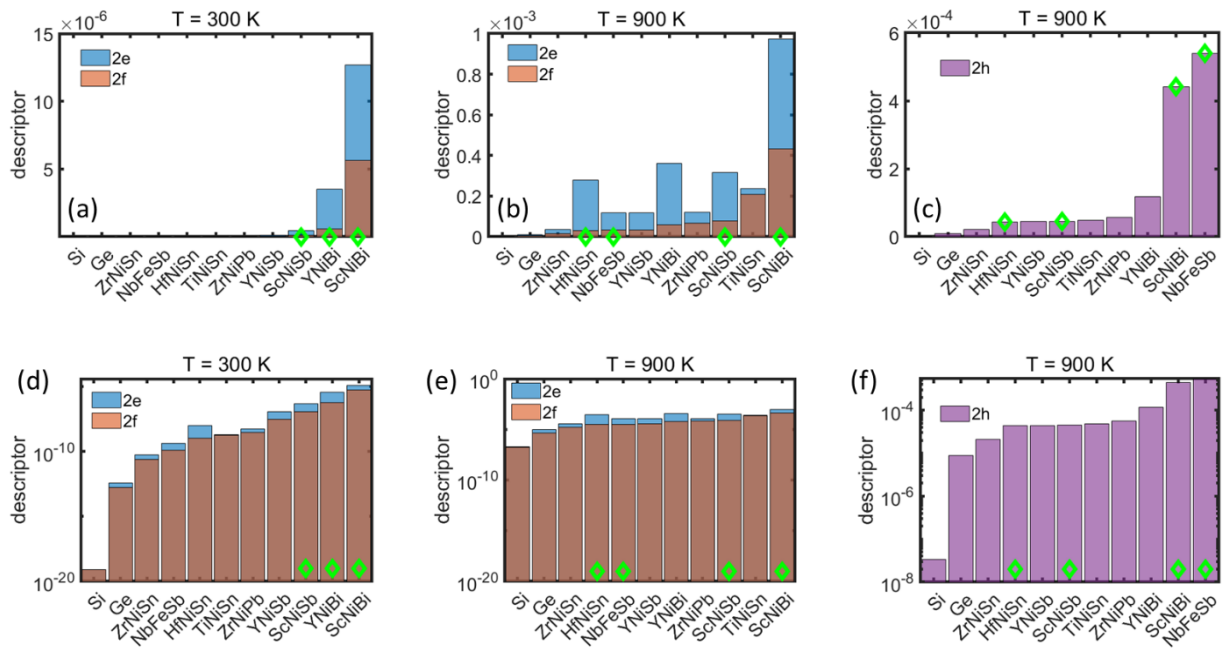

Figure S3: Material descriptors. (a) Descriptors 2d and 2f for 300 K. (b) Descriptors 2d and 2f for 900 K. (c) Descriptor 2h for 900 K. (a-c) linear scale. (d-f) logarithmic scale.

We underline that the presented descriptor aims at ranking the PF peak in the intrinsic regime, regardless of if it is higher than the common PF peak. In the main paper Figure 2, we discussed only the materials where the additional high PF peak in the intrinsic region is much higher than the conventional PF peak, for example YNiBi and TiNiSn, show this peak in the pristine region, but its value is lower than the conventional peak in the high doped regime and have not been presented in the manuscript.